\documentclass[a4paper,twocolumn,english,aps,prb,superscriptaddress,showpacs,showkeys]{revtex4}
\usepackage[T1]{fontenc}
\usepackage[latin9]{inputenc}
\usepackage{amsmath}
\usepackage{graphicx}
\usepackage{subfigure}
\usepackage{hyperref}
\usepackage{babel}
\usepackage{color}

\begin{document}
\date{\today}

\title{Magnetism- and impurity-assisted chain creation in Ir and Pt break junctions}
\author{Solange Di Napoli}
\affiliation{Departamento de F\'{\i}sica de la Materia Condensada, CAC-CNEA, Avenida
General Paz 1499, (1650) San Mart\'{\i}n, Pcia. de Buenos Aires, Argentina}
\affiliation{Consejo Nacional de Investigaciones Cient\'{\i}ficas y Técnicas,
CONICET, Buenos Aires, Argentina}

\author{Alexander Thiess}
\affiliation{German Research School for Simulation Sciences, D-52425, J\"ulich, Germany}
\affiliation{ Peter Gr\"unberg Institut and Institute for Advanced Simulation, Forschungszentrum J\"ulich and JARA, 52425 J\"ulich, Germany}

\author{Stefan Bl\"ugel}
\author{Yuriy Mokrousov}
\affiliation{ Peter Gr\"unberg Institut and Institute for Advanced Simulation, Forschungszentrum J\"ulich
 and JARA, 52425 J\"ulich, Germany}

\date{\today}
\begin{abstract}
Applying the generalization of the model for chain formation in break-junctions
{[}JPCM \textbf{24}, 135501 (2012){]}, we study the effect of light
impurities on the energetics and elongation properties of Pt and Ir chains. Our model enables 
us with a tool ideal for detailed analysis of impurity assisted chain formation, where zigzag bonds play an
important role. In particular we focus on H (\textit{s}-like) and O (\textit{p}-like)
impurities and assume, for simplicity, that the presence of impurity
atoms in experiments results in ..M-X-M-X-... (M: metal, X: impurity)
chain structure in between the metallic leads. Feeding our model with material-specific parameters 
from systematic full-potential  first-principles calculations, 
we find that the presence of such impurities strongly affects the binding properties
of the chains. We find that while both types of impurities enhance the probability of chains to
be elongated, the $s$-like impurities lower the chain's stability. We also analyze the effect 
of magnetism and spin-orbit interaction on the growth properties of the chains.
\end{abstract}
\maketitle

\section{Introduction}
The elongation of some metal junctions allows the fabrication of the
self-supported one-dimensional systems: the suspended atomic chains, 
achieved, for example, in a mechanically controllable break
junction (MCBJ)~\cite{Ruitenbeek01}. In such an experiment a metallic
junction which has been created by lithography is being stretched
until it breaks. If this stretching is performed in a controllable
way by using piezoelements then the creation of stable monowires can
be observed. With this technique it has been possible to create monoatomic
chains from Ir~\cite{Ryu06}, Pt~\cite{Ruitenbeek01} and Au~\cite{Ohnishi98,Ruitenbeek98,Rubio01}
up to a length of 5 to 6 atoms for Pt and even more for Au which has
been confirmed by measuring the conductance during the formation.
One of the most exciting perspectives in this area is the ability
to study not only theoretically but also experimentally some of
the most fundamental properties of one-dimensional systems, which are
mostly related to their transport properties. As magnetism is enhanced in 
such systems, the latter are promising candidates for spintronics applications
due to a possibility to simultaneously probe, control and switch the magnetic 
state by spin-polarized  electrical currents~\cite{Heinrich06,Tosatti09,Calvo09,Lounis13}.

Since the creation of the first free-standing atomic chains of
gold atoms in 1998~\cite{Ruitenbeek01,Ohnishi98}, a lot of work
has been done in search for other elements that could also form
atomic chains. Although there were some reports on the formation of several
$3d$- and $4d$-row metallic chains, it has not been definitely established
whether it is possible to successfully create long atomic chains out of
other elements than Au, Pt and Ir. One of the crucial points in the success of chain
formation is the reconstruction of the low-index surfaces of the selected
elements: gold nanowires spontaneously evolve into freely suspended
chains of atoms. R.H.M. Smit \textit{et al.}~\cite{Ruitenbeek01}
showed that all the $5d$ metals that show similar reconstructions
(Ir, Pt and Au) also form chains of atoms.

In the last few years it has been possible to strengthen the bonds in a suspended
chain and achieve a higher probability of chain producibility by adding external 
absorbates during the chain formation process. It is known that low-coordinated atoms are chemically 
more reactive than in bulk~\cite{Barnett04}, thus, being the coordination number only of two in an atomic chain, 
chains are expected to be even more reactive than nanoparticles, opening the possibility
for molecular absorbates to dissociate, even at low temperatures. For instance, O atoms are 
expected to be incorporated in the chains, as predicted in several previous 
works~\cite{Novaes06,Bahn02,Thijssen06}.
Oxygen atoms are not the only kind of impurities that can help
in the process of chain formation. There are previous reports
on the effect of H, B, C, N and S impurities on gold chains~\cite{Novaes03,Barnett04,Ugarte02}.
All of them found that the inserted atoms in gold nanowires form not
only stable but also very strong bonds. 
In a previous work~\cite{Dinapoli12}, we systematically applied our chain formation model for a detailed
study of the trends in the formation of Cu, Ag and Au chains in pure break junctions as well as Cu, Ag and
Au break junctions with the atmosphere contaminated by H, C, N and O impurities. We also extended the chain
formation model to the case of a geometrically more complex planar zigzag arrangements of the atoms in the
chain. We have demonstrated that adding the mentioned environmental impurities 
in the noble-metal chain formation process leads to a significantly enhanced probability for chain formation 
in noble metals break junctions. In this work we systematically apply our chain formation model in
late $5d$ transition metals (TMs), namely, Pt and Ir. We consider linear and zigzag arrangements and two different types of 
impurities, namely $s$-like (H) and $p$-like (O). We include spin polarization in all the calculations and
carefully check the effect of spin-orbit coupling on our findings. Overall, we find that in analogy to noble chains
the presence of impurities enhances the probablity of chains to form. And while $p$-like impurities also
strengthen the stability of the chains, the $s$-like impurities enhance the spin moments of the transition
metals and lower their structural stability, which is largely dependent on the directional bonding between
the atoms due to presence of the $d$-states at the Fermi energy. 

The paper is organized as follows. In the next section a brief
description of the computational details used in the calculations
is given. In Section \ref{results} we present the calculated
structural and magnetic properties of the linear and zigzag chains, for both single-atom and chains with impurities.
In Section \ref{P-model} the producibility (P-)model
is breafly recalled and the results are presented. In this section we also analyze the role of magnetism
and spin orbit coupling in the chain formation process. Finally, we present the conclusions in 
Section \ref{Conclusions}.

\section{Computational details}
\label{comp-det} 
\begin{figure}
\begin{centering}
\includegraphics[width=1.0\columnwidth]{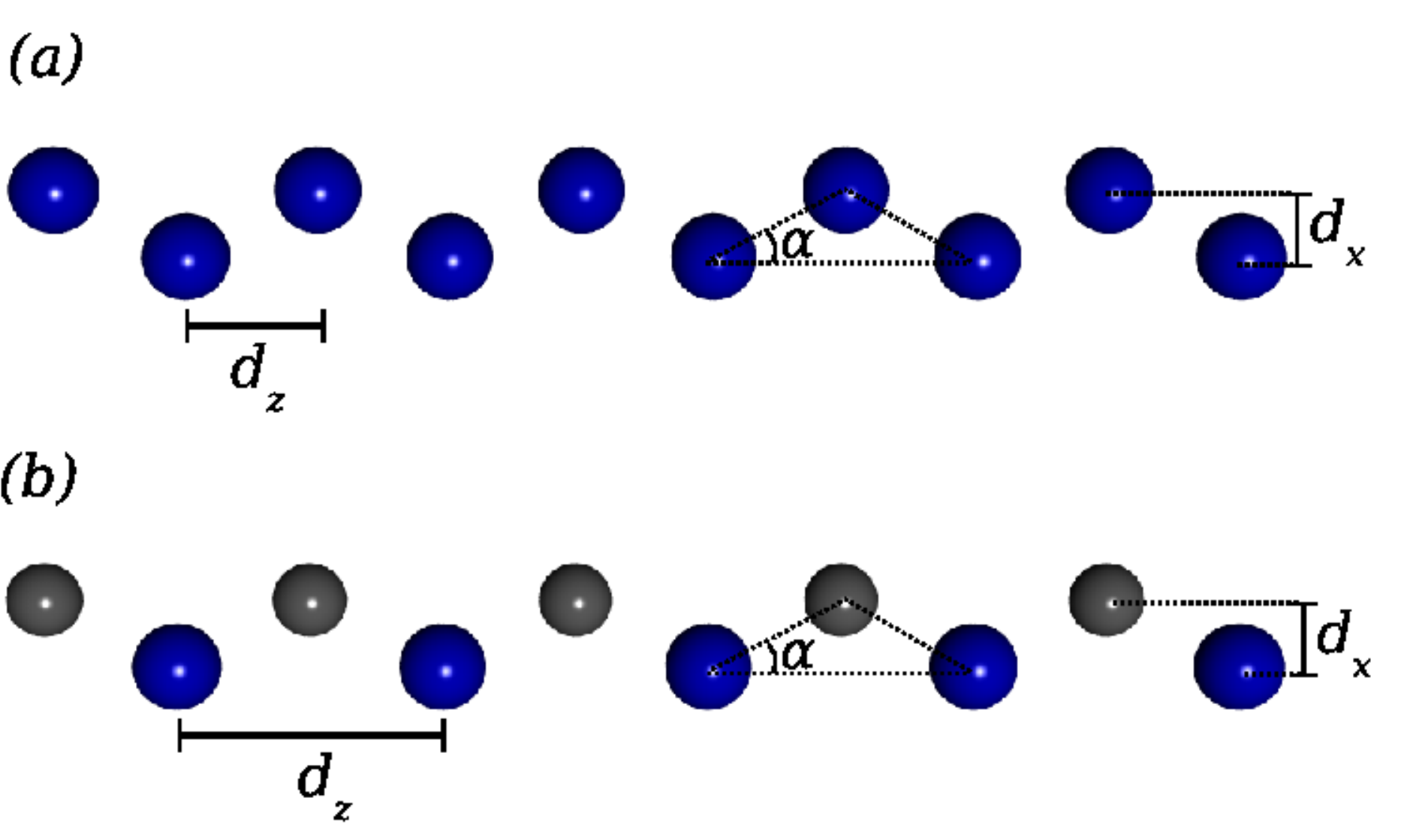}
\par\end{centering}
\caption{(Color online) Schematic structure for (a) single-atom planar zigzag
chains and (b) planar zigzag chains with impurities. The infinite zigzag chains make an $\alpha$-angle
with the chain axis, $z$. $d_{z}$ is the projection of the noble-metal to noble-metal distance
into the chain axis and $d_{x}$ is the projection into a perpendicular degenerate axis. The metal atoms are 
represented by large blue spheres, while smaller gray spheres stand for the impurity atoms.}
\label{definitions} 
\end{figure}

In the present first principles calculations, we
employ the full-potential linearized augmented plane-wave method
for one-dimensional (1D) systems~\cite{Yura1D}, as implemented
in the FLEUR code~\cite{FLEUR}. The calculations are based on density
functional theory within the (rev-PBE) generalized gradient approximation (GGA)
to the exchange-correlation potential.
Basis functions were expanded up to $k_{max}=$4.0~bohr$^{-1}$ and
we have used 32 $k$ points in one half of the 1D Brillouin zone.
The position of the boundary between the interstitial and the vacuum
region $D_{vac}$ as well as the in-plane lattice constant $\tilde{D}$,
used for the generation of reciprocal lattice vectors were set to 8.3~bohr
and 9.9~bohr, respectively. For details of the method and spatial partitioning within the
1D FLAPW scheme see Ref.~\onlinecite{Yura1D}. The muffin-tin radii have been set to $R_{MT}=$1.90~bohr
for Pt and Ir (with local orbitals in the $5s$ and $5p$ states),
and $R_{MT}=$1.0~bohr for the impurities. These values were chosen not only to guarantee the
good convergence of the obtained results but also to achieve a wide range of zigzag $\alpha$-angles
(see Fig.~\ref{definitions}). For our calculations we included the effect of the spin-orbit
interaction, and considered the magnetization direction in the three directions of space ($x$, $y$ and $z$),
as none of them are equivalent when the zigzag geometry is taken into account.

We have set the coordinate system such that the chains are aligned
along the $z$ axis, and considered a two-atoms unit cell to allow for
zigzag arrangements. In Fig.~\ref{definitions} a schematic
picture of the infinite planar zigzag atomic chains
is shown, where the defined distances ($d_{z}$, $d_{x}$) as well as the zigzag $\alpha$-angle are presented.

\section{Structure and magnetic properties}
\label{results}

\begin{figure}
\begin{centering}
\includegraphics[width=1.0\columnwidth]{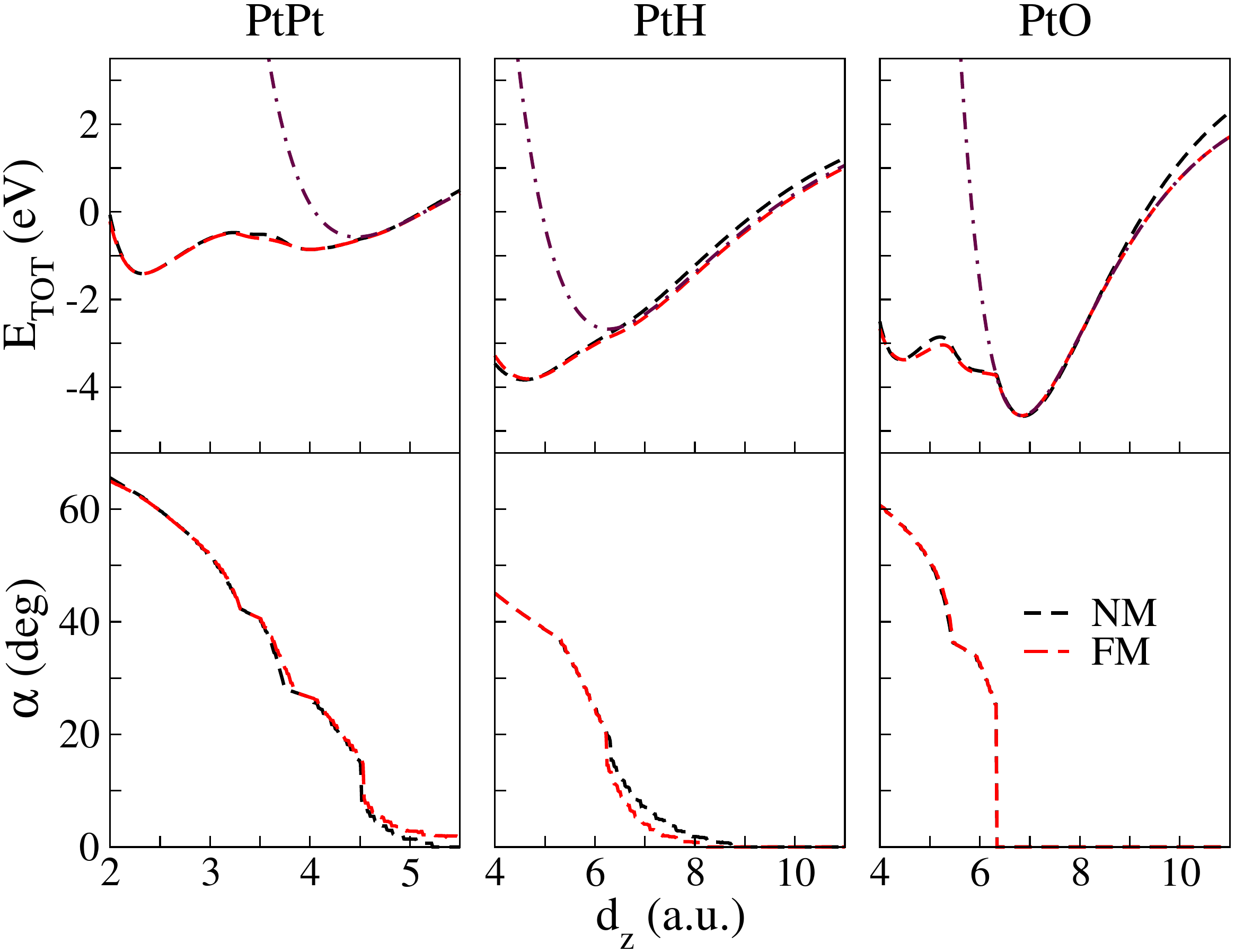}
\end{centering}
\caption{(Color online) Upper panel: the minimal total energy $E_{TOT}$ (per atom) as a function of the interatomic
distance ($d_{z}$) for PtPt (left), PtH (middle) and PtO (right). Results of non-magnetic (NM) and ferromagnetic (FM) calculations are presented.
The dashed-dotted maroon lines correspond to the total energies of the linear infinite FM chains.
Lower panel: the evolution of the corresponding zigzag $\alpha$-angles.}
\label{Pt} 
\end{figure}

\begin{figure}
\begin{centering}
\includegraphics[width=1.0\columnwidth]{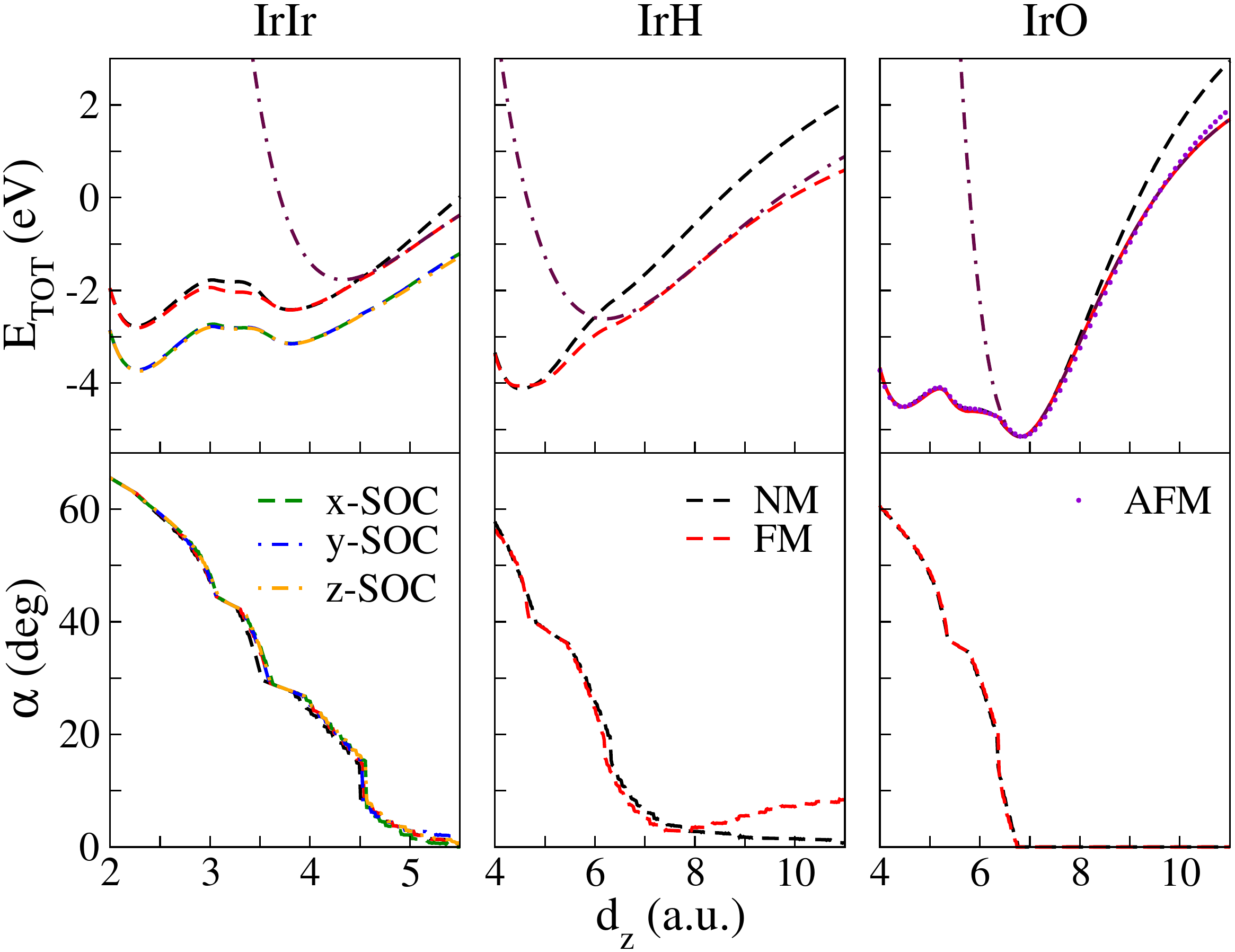}
\end{centering}
\caption{(Color online) Upper panel: the minimal total energy $E_{TOT}$ (per atom) as a function of the interatomic
distance ($d_{z}$) for IrIr (left), IrH (middle) and IrO (right). Results of non-magnetic (NM) and ferromagnetic (FM) calculations are presented.
For IrIr chains, results when applying spin-orbit coupling are also shown.  
The dashed-dotted maroon lines correspond to the total energies of the linear infinite FM chains.
Lower panel: the evolution of the corresponding zigzag $\alpha$-angles.}
\label{Ir} 
\end{figure}

As one of the main goals of this work is to understand the incidence of the geometric structure as well as 
magnetism and spin-orbit interaction in chain formation process from energetic considerations, we have 
performed systematic calculations of linear and zigzag geometries for two late $5d$ TMs, Ir and Pt, with
$s$-like (H) and $p$-like (O) impurities. We carried out total energy calculations as a function of both interatomic
parameters, $d_z$ and $d_x$, as defined in Fig.~\ref{definitions}. For each of the five chosen $d_x$ distances 
we performed self-consistent calculations for at least ten different $d_z$ values, fitted each energy curve with a Morse-potential
and interpolated the two-dimensional energy profile in the $d_z$-$d_x$ plane, from which we extracted the 
effective minimum energy curves as a function of $d_z$, $\epsilon(d_z)=min_{d_x}\epsilon(d_z,d_x)$.
Our main results for single-atom Pt and Ir chains are summarized in the left panels of Figs.~\ref{Pt} and \ref{Ir}, respectively, where the total energies
and the zigzag $\alpha$-angles are plotted as a function of $d_z$. 
We obtain the minimum energy curves for different magnetic configurations, namely non-magnetic (NM), ferromagnetic
(FM) and including spin-orbit coupling with the magnetization pointing along three directions of space, namely x-SOC, y-SOC and z-SOC. We present,
in Figs.~\ref{Pt} and \ref{Ir}, only a selection of the collected data that ilustrates the main results, to avoid confusing the reader with too many curves.
Total energies of the FM Pt and Ir infinite linear-chains are also shown for comparison.  We find a characteristic two-well structure in the energy curves for both
 Pt and Ir, where the first minima are located around
$62^{\circ}$ for both atomic chains, and the second minima are around $27^{\circ}$ in the Pt case and around $28^{\circ}$ 
in the Ir chains, showing that these metals prefer higher coordination and closely packed structures. Our results agree well with those reported by 
Fern\'andez-Seivane\cite{Ferrer07, Ferrer09} and by Tung,\cite{Tung10} except for the location of the second minimum at $\alpha\sim 45^{\circ}$ in Ir chains found
by Fern\'andez-Seivane.  
It is interesting to note that inclusion of the spin-orbit interaction does not modify the general trends of the energy profiles in none of the studied systems.

To study the effect of environmental impurities in Pt and Ir chains we considered two prototypes: $s$-like (H), and $p$-like (O) impurities.
We performed non-magnetic and spin-polarized calculations in the scalar-relativistic (FM) and fully relativistic (SOC) approximations. 
A collection of representative results is presented in the middle and right panels of Figs.~\ref{Pt} and \ref{Ir}, respectively, where we show the total energy 
curves as well as the evolution of the $\alpha$-angles as a function of $d_z$, for Pt-X and Ir-X (X=H,O) chains, where
we also included the energies of the corresponding infinite linear chains. We find that, when including SOC, the enegies are lower than the corresponding FM energies, but 
the curve's features are similar, giving the same structural properties. 
The trends observed in the energetics of the NM results are similar to those obtained from the FM calculations. 
One feature that has to be pointed out is that in general we obtain that both Pt and Ir, not only in 
single-atom chains but also with impurities, are ferromagnetic, as can be seen from the uper panels of the figures. 
This property is enhanced when stretching the atomic chains. Nevertheless,
there is a range of $d_z$ values where we obtain that NM structure is more stable than the FM one, when the impurity is $s$-like. 
This range is smaller for  Pt than for Ir.
With respect to the geometric structure, we find that when including an $s$-like impurity the planar zigzag structure is more stable,
with $\alpha \sim 40^\circ$ for both Ir and Pt chains. The energy curves in the vicinity of this angle are rather flat.
When we introduce a $p$-like impurity the energy curves are quite different as the ones obtained for the single-atom chains, as the two-well structure is no longer 
present and, instead, a three minima curve emerges. The global minima are located in $d_z$ 
corresponding to linear chains, indicating that the presence of this kind of impurities leads to an effective straightening of the bonds in the chains. 

A common feature for all Pt-based chains, which can be seen from the comparison between Figs.~\ref{Pt} and \ref{Ir} is that their energy-curves have smaller
slopes that the ones displayed by the Ir-based chains. This softening of the energy profile can be related to the fact that the $d_{xy}$ and $d_{x^2-y^2}$ states of Pt
are below the Fermi energy ($E_F$) whereas in the case of Ir these states are crossing $E_F$, thus providing an additional channel for hybridization and bonding between 
the atoms. A similar behavior was also found when comparing Au and Ag chains~\cite{Dinapoli12}. This fact has a high impact on the chain stabilization process, as we 
will discuss in the next section.   

For IrO chains we have also tested the possibility of the antiferromagnetic (AFM) solution  (see Fig.~\ref{Ir}). We find that there are two ranges of $d_z$ values
 where the AFM solution is more stable, namely, for large $\alpha$ values ($4.0\mathrm{bohr}<d_z<4.4\mathrm{bohr}$) and when $d_z \in$ ($6.85\mathrm{bohr}<d_z<9.55\mathrm{bohr}$). 
Concerning the geometric structure derived from the features of the energy profiles, there
are no substantial differences between the FM and AFM solutions and, therefore, we do not consider this magnetic structure when analyzing the chain formation
process in the following section.
Summarizing, we find that the structural properties are essentially independent on magnetism or presence of 
the spin-orbit interaction.

\begin{figure}
\noindent \begin{centering}
\includegraphics[width=1.0\columnwidth]{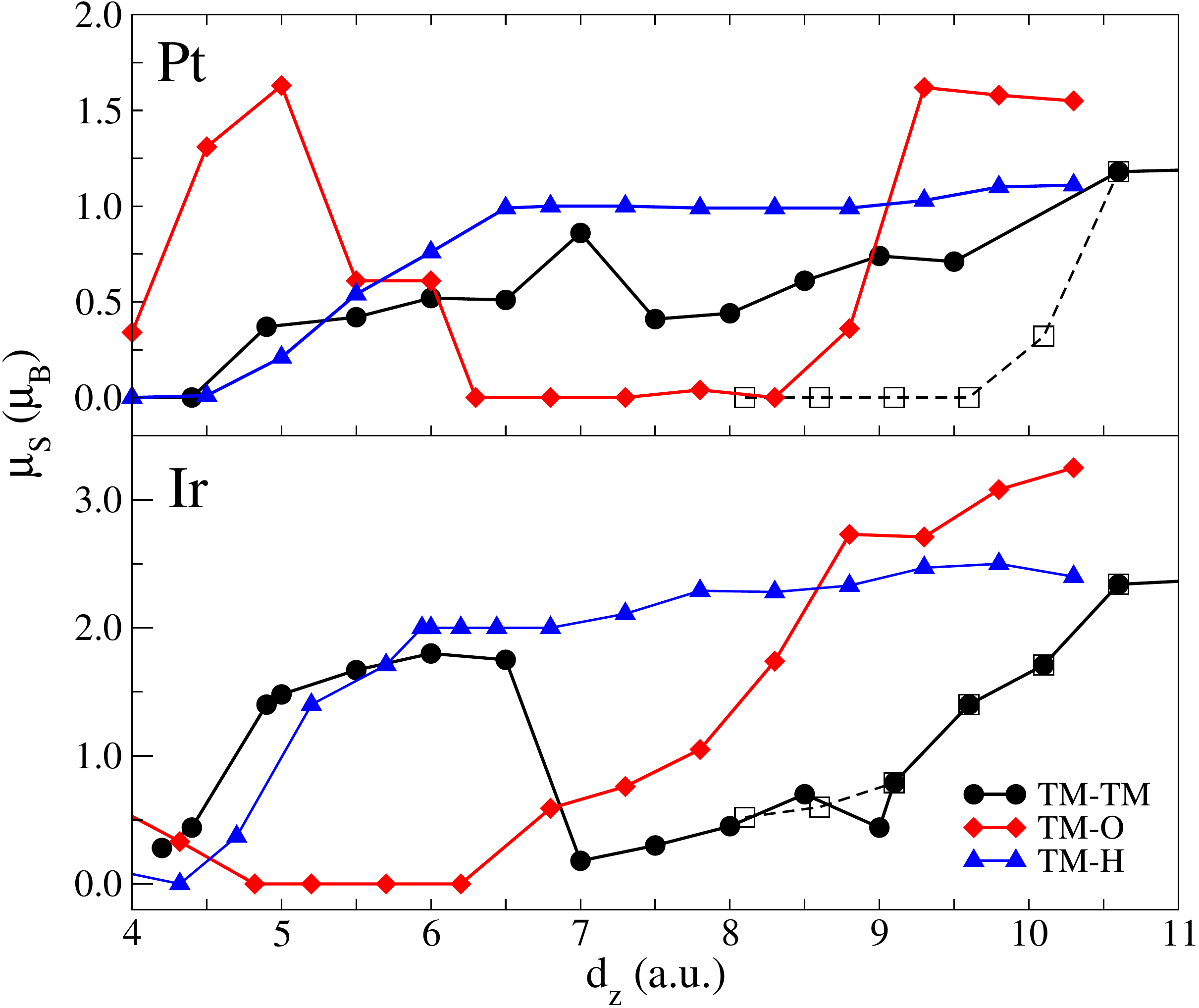}
\par\end{centering}
\caption{(Color online) Atomic magnetic moments of the metallic atoms (in $\mu_{\mathrm{B}}$)  as a function of $d_{z}$ in Pt, Pt-X (upper panel), Ir and Ir-X (lower panel) chains.
Open black squares correspond to Pt (Ir) linear chains, black circles to Pt (Ir) zigzag chains, red diamonds to Pt (Ir) with O-impurities and blue triangles to Pt (Ir) with
H-impurities.  In the case of the monowires (both linear and zigzag chains), the double cell is taken into account, to show the results in the same $d_z$-range.}
\label{MM-X_moments}
\end{figure}

We turn now our attention to the magnetic properties of the studied chains. In this work, we do not report on magnetic anisotropy enegies (MAEs) and orbital moments,
as these two quantities for Pt and Ir chains were extensively studied in the past~\cite{Tosatti03,Ferrer07,Weht08,Ferrer09,Tung10,Alex10}. The evolution of the atomic spin 
moments with $d_z$ is presented in Fig.~\ref{MM-X_moments}. In the upper panel we show the spin moments for the Pt-Pt and
Pt-X (X=H,O) cases while in the lower panel the Ir-Ir and Ir-X moments are plotted. In case of impurities, the metal's magnetic moments are calculated as the unit cell moment minus the local moment of the X atom, 
as the muffin-tin radii we use are too small to give reasonable values of the magnetic moments of the metal atoms but are big enough to describe the local moment on the X atom. 
As it can be seen from this figure, magnetic moments increase almost monotonically in Pt chains, whereas in the case of Ir chains they show a high-spin to low-spin to high-spin
transition in the range of $d_{z}$ which corresponds to the formation of the zigzag minimum during the stretching. Both cases are magnetic in their ground state zigzag structures, and their 
magnetic moments at corresponding distances are similar, $\mu_{S}\sim0.44 \mu_{\mathrm{B}}$. All our calculated magnetic moments agree well with the ones calculated by 
Fernández-Seivane \textit{et al.}\cite{Tosatti03,Ferrer07,Ferrer09,Tung10}

Both $5d$ TMs, when combined with H show a monotonous increment in their magnetic moments from zero to the saturated value, the one obtained for the monoatomic wires. 
Including O in the atomic chains leads to an interesting magnetic behavior. In the Pt case, a transition between high-spin to low-spin to high-spin is present, in contrast to 
what was observed in the Pt monowire, and the same trend is shown by the O atoms (not shown). In the Ir case, while the high-spin to low-spin transitions were present in the 
Ir monowires, they are no longer present when combining Ir with O. In this case, the Ir magnetic moments are negligible until the distance at which a growth up to huge values begins. The O moments show the same trend, with smaller magnetic moments.

\begin{figure}[ht!]
\begin{center}
\includegraphics[width=1.0\columnwidth]{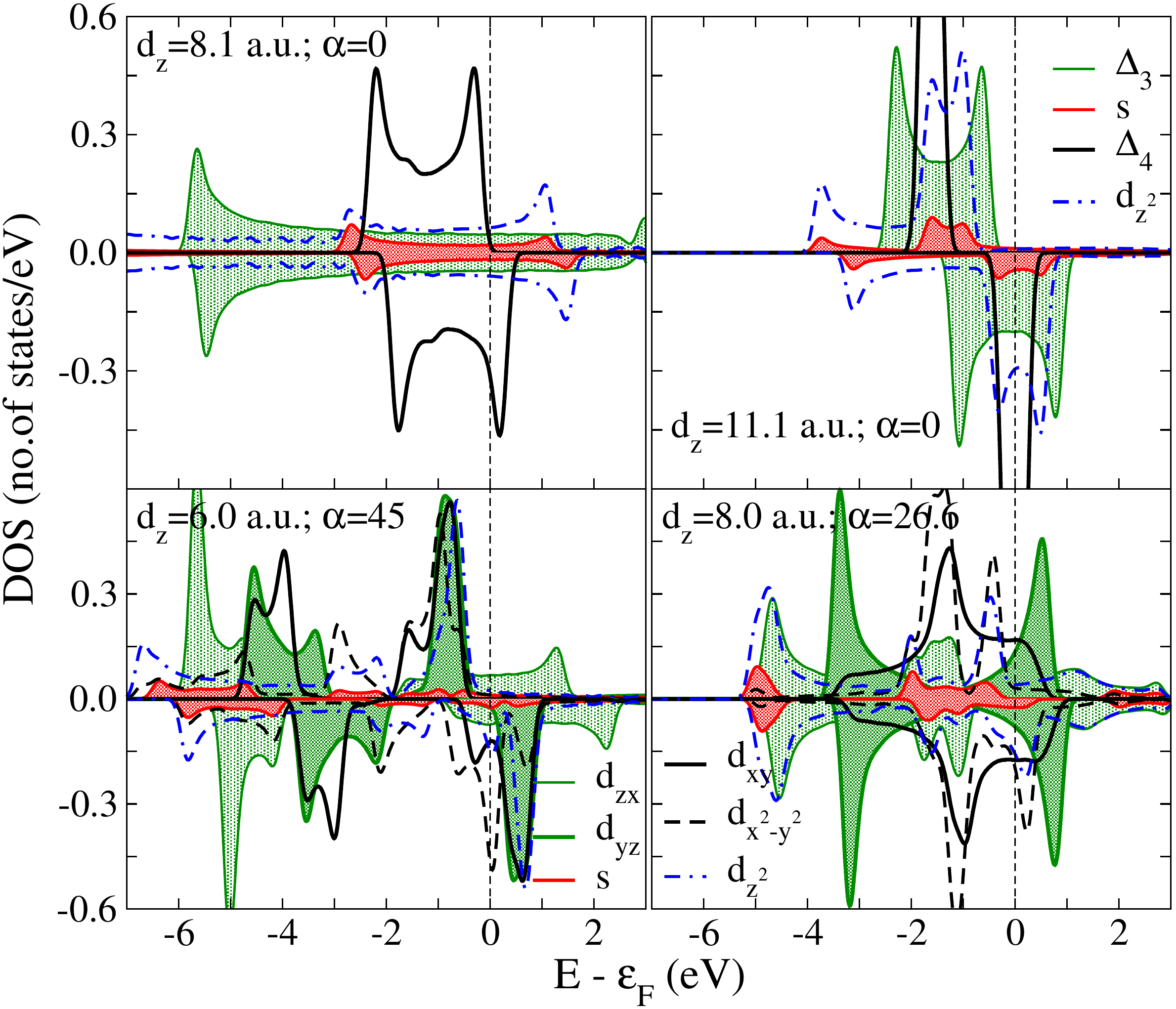}
\caption{(Color online) Orbital decomposition of the density of states of Ir chains in selected geometries. The vertical black line represents the Fermi 
energy ($\epsilon_F$).}
\end{center}
\label{dos}
\end{figure}

An example of how the ferromagnetic density of states (DOS) evolves as the parameter $d_{z}$ increases is shown in Fig.~5, in which we refer to some selected cases. 
In the upper panels the DOS of infinite linear chains corresponding to small and large Ir-Ir distance is presented. As can be seen, the different orbitals in the small
distance case are very extended and the exchange splitting is negligible, giving rise to zero magnetic moment. When the interatomic distance in the monowires is increased, 
the tendency to ferromagnetic order and the exchange splitting are strongly enhanced giving rise to huge magnetic moments (see lower panel of Fig.~\ref{MM-X_moments}). The 
cylindrincal symmetry present in the linear chains is also verified in the upper panels of Fig.~5, where the $d_{xy}$ and $d_{x^{2}-y^{2}}$($\Delta_{4}$-symmetry) orbitals
are degenerate, as well as the $d_{zx}$ and $d_{yz}$ ($\Delta_{3}$-symmetry) orbitals.  These degeneracies are broken when the chain has a zigzag geometry, as can be observed in the 
DOS of two different zigzag $\alpha$-angles presented in the lower panels of Fig.~5, corresponding to a low-spin ($d_{z}=8\,$bohr; $\alpha=26.6^{\circ}$) and a 
high-spin ($d_{z}=6\,$bohr; $\alpha=45^{\circ}$). 
In the high-spin case (lower left) the orbitals are more extended, due to a smaller Ir-Ir bonding distance and the exchange 
splitting is large, consistent with high magnetic moments. On the other hand, in the low-spin case (lower right) the bands are narrower due to a larger Ir-Ir bonding distance, 
and the exchange splitting is small but not negligible, giving rise to a small magnetic moment.

\section{Producibility and stability of the imputity-assisted chains}
\label{P-model}

In this work, we apply the generalization of our model for chain formation in break-junctions~\cite{Dinapoli12}, 
based on total energy arguments where the zig-zag and linear geometries as well as the presence of impurities are taken into account.
To breafly recall this model, we divide the system into two regions: the leads and the suspended chain. The electronic structure of the two parts is
considered separately, thus neglecting their mutual influence. This approximation is good enough to describe the formation of long chains [X], which is the purpose  of the current work. Within our model, the whole process of chain formation consists
of two consequent processes: first, one atom is extracted out of the lead into the chain, reducing the coordination of this particular
atom. Therefore, additional external energy is necessary, which we account for by the difference of the cohesion energy for an atom in the lead, $E_{lead}$, and in the chain, $E_W(d_0)$, 
both at equilibrium distance, thus giving $\Delta E_{lead}=E_W(d_0)-E_{lead}$. 

The second process is related to the relaxation of all the chain bonds to a smaller interatomic distance after the additional atom has entered the chain. In this way,
the chain formation process can be translated into the following equation (P-criterion):
\begin{equation}
E^*(d,N)=(N+1) \epsilon (d) - [\Delta E_{lead} - (N+2) \epsilon (\tilde{d})]
\label{E*}
\end{equation}
where $d=L/(N+1)$ and $\tilde{d}=L/(N+2)$ are the chain interatomic distances before and after the elongation, respectively, $N+1$ is the number of bonds in a chain of 
$N$ atoms and $L$ is the distance between the leads. 
The binding energy of the suspended chain, $\epsilon(d)$, is defined from the binding energy of the infinite wire with interatomic distance $d$, $E_W(d)=E_W(d_0)+\epsilon(d)$, 
(we always have two atoms in the unit cell), relative to the wire's cohesive energy at equilibrium interatomic distance $d_0$, $E_W(d_0)$, which implies $\epsilon(d)\textgreater 0$.

The chain formation energy is given by $\Delta E_{lead}=E_W(d_0)-E_{lead}$, where $E_{lead}$ denotes the cohesive energy of the lead atom. 
$E_{lead}$ is calculated as explained in Ref.~\onlinecite{Dinapoli12} and the values of the energy barriers we use for Pt-based chains are $\Delta E_{lead}=$1.852~eV, 1.791~eV 
and 1.900~eV  for NM, FM and SOC calculations, respectively. The corresponding ones for Ir-based chains are $\Delta E_{lead}=$2.809~eV, 2.684~eV and 2.693~eV. It is worthy saying that here we assume that the light impurities are present in the atmosphere and not bound at the surface. This assumption was previously well 
justified.\cite{Bahn02,Dinapoli12}   
 
\begin{figure*}[ht!]
\begin{center}
\begin{tabular}{ccc}
\includegraphics[width=3.00cm, angle=270]{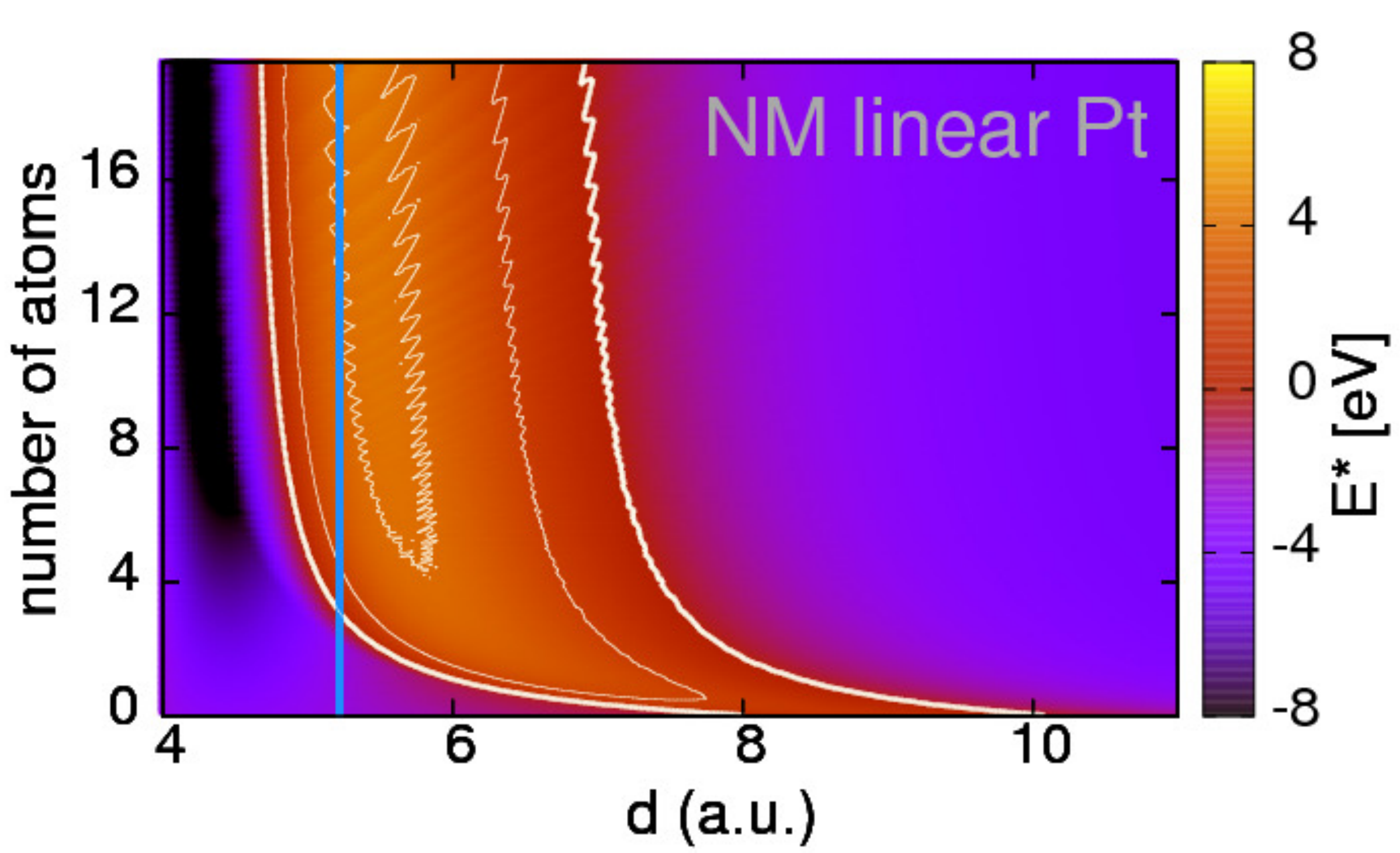} & & 
\includegraphics[width=3.00cm, angle=270]{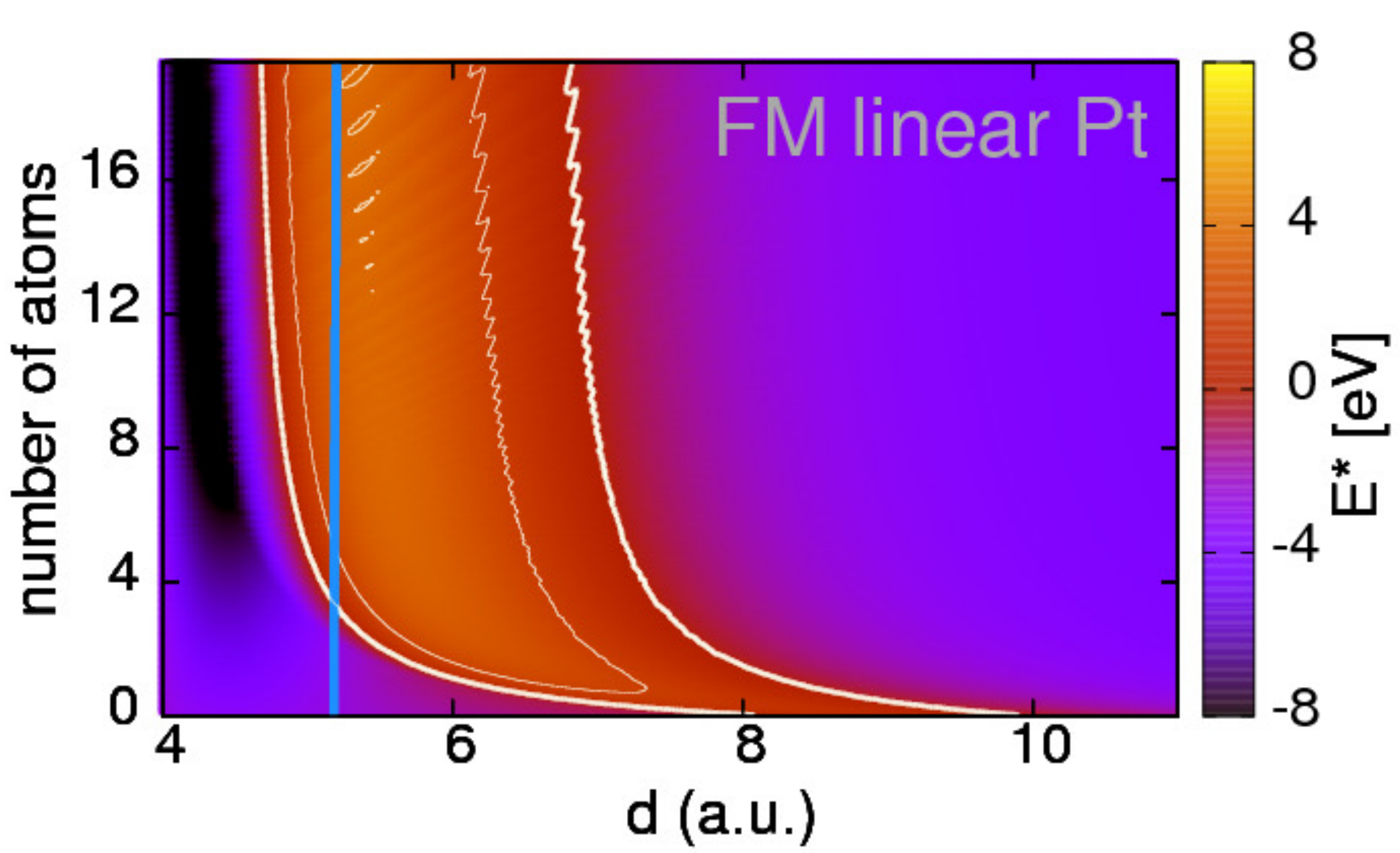}\\ 
\includegraphics[width=3.00cm, angle=270]{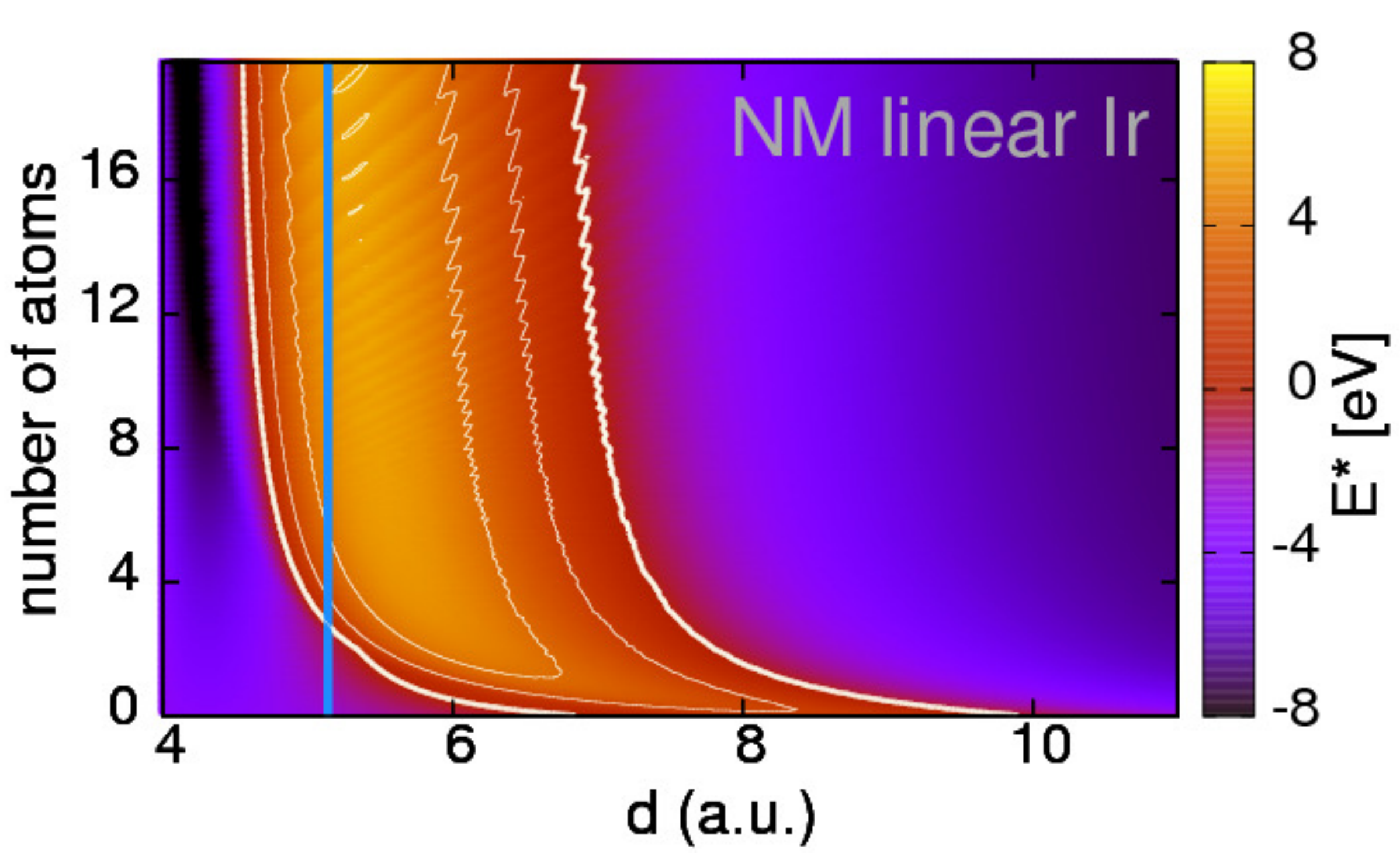} & & 
\includegraphics[width=3.00cm, angle=270]{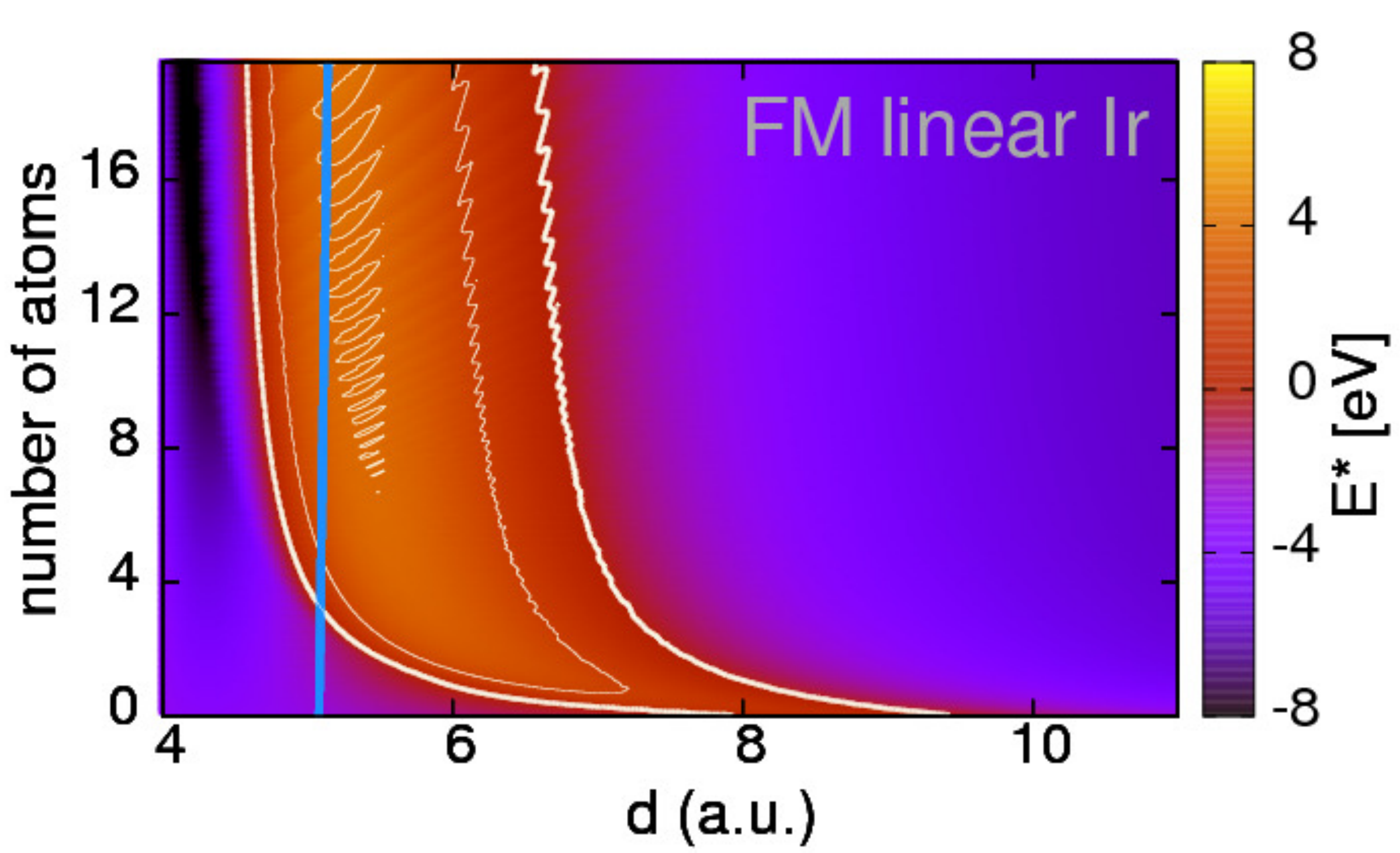}\\ 
\includegraphics[width=3.00cm, angle=270]{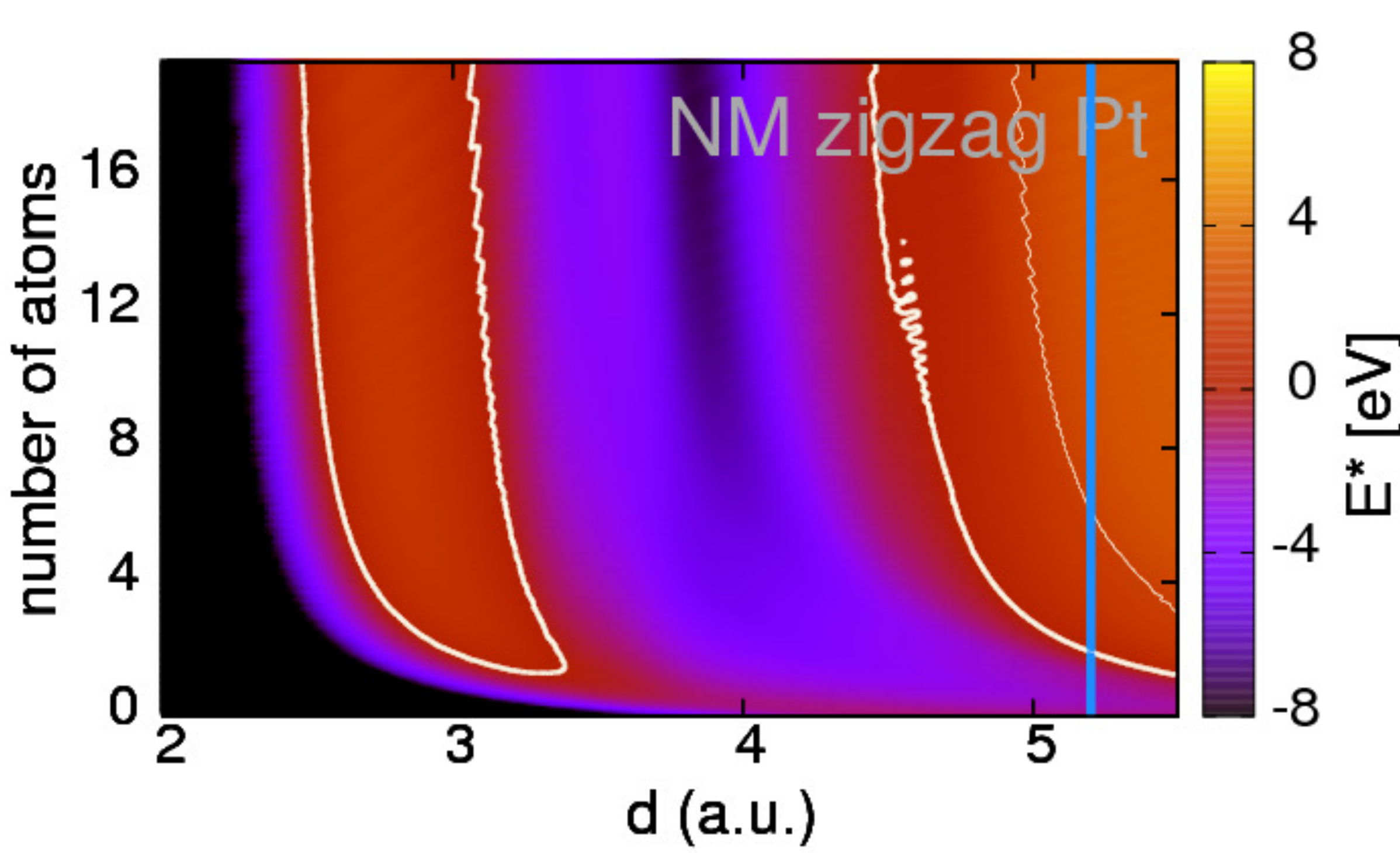} & & 
\includegraphics[width=3.00cm, angle=270]{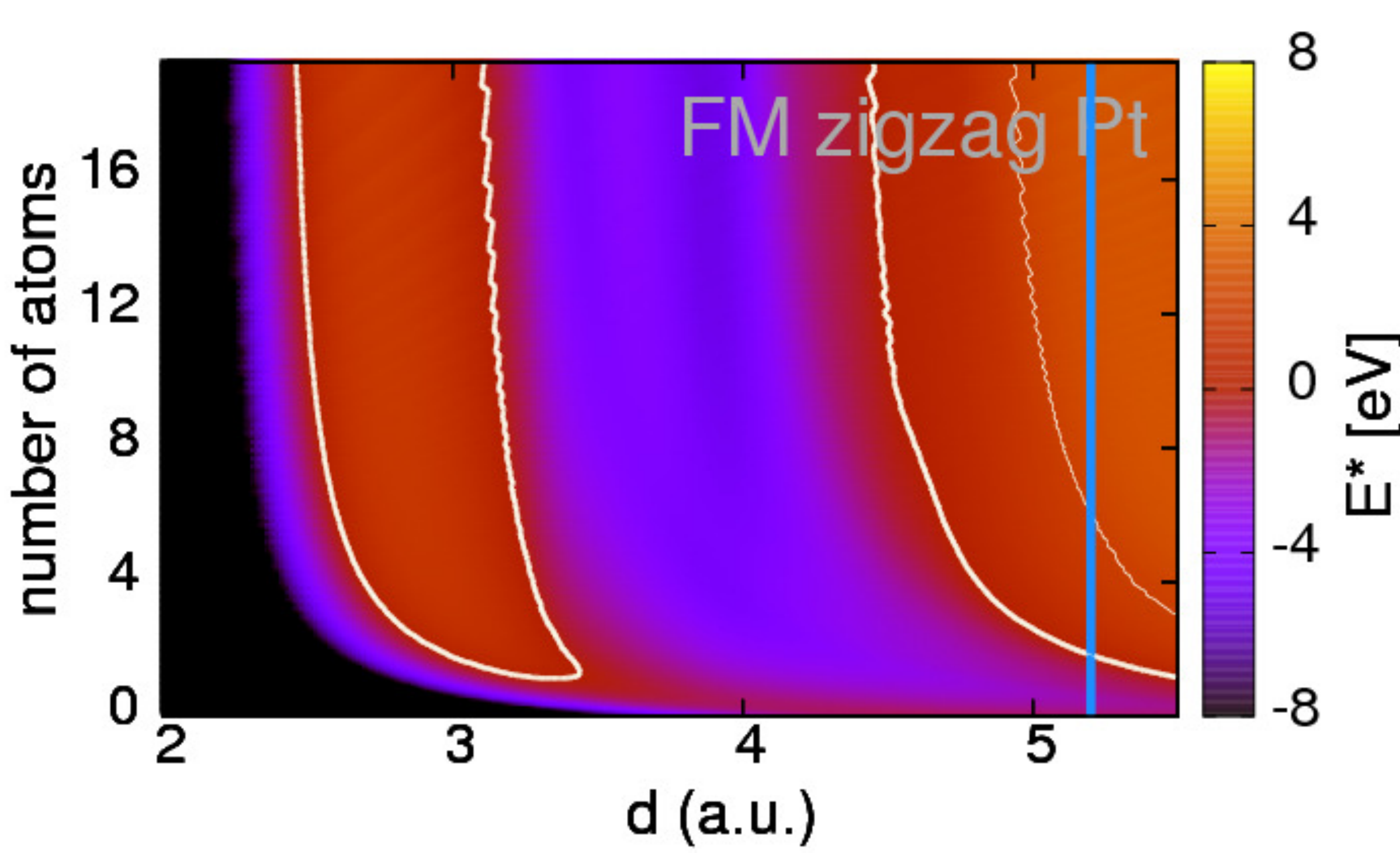} \\
\includegraphics[width=3.00cm, angle=270]{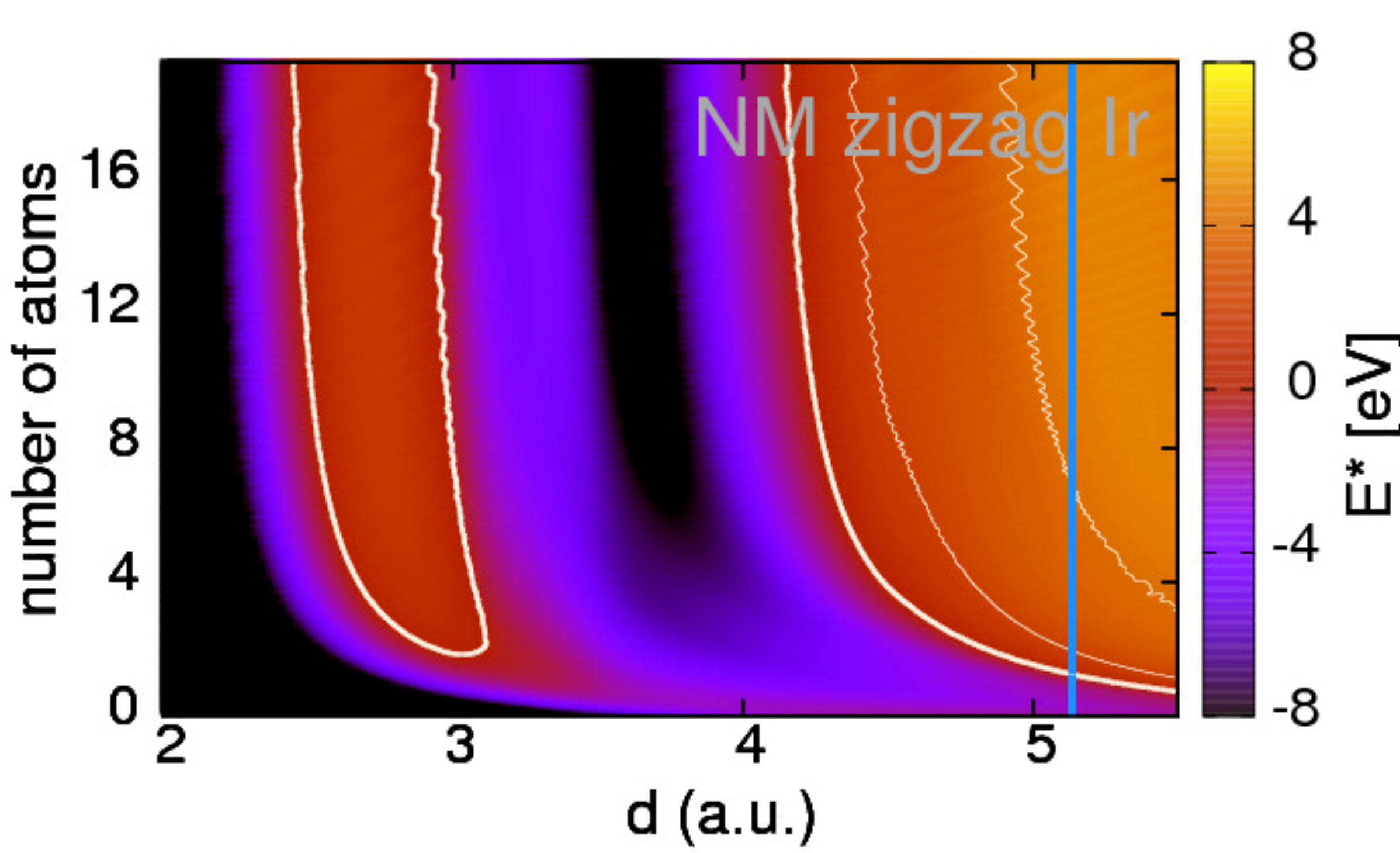} & & 
\includegraphics[width=3.00cm, angle=270]{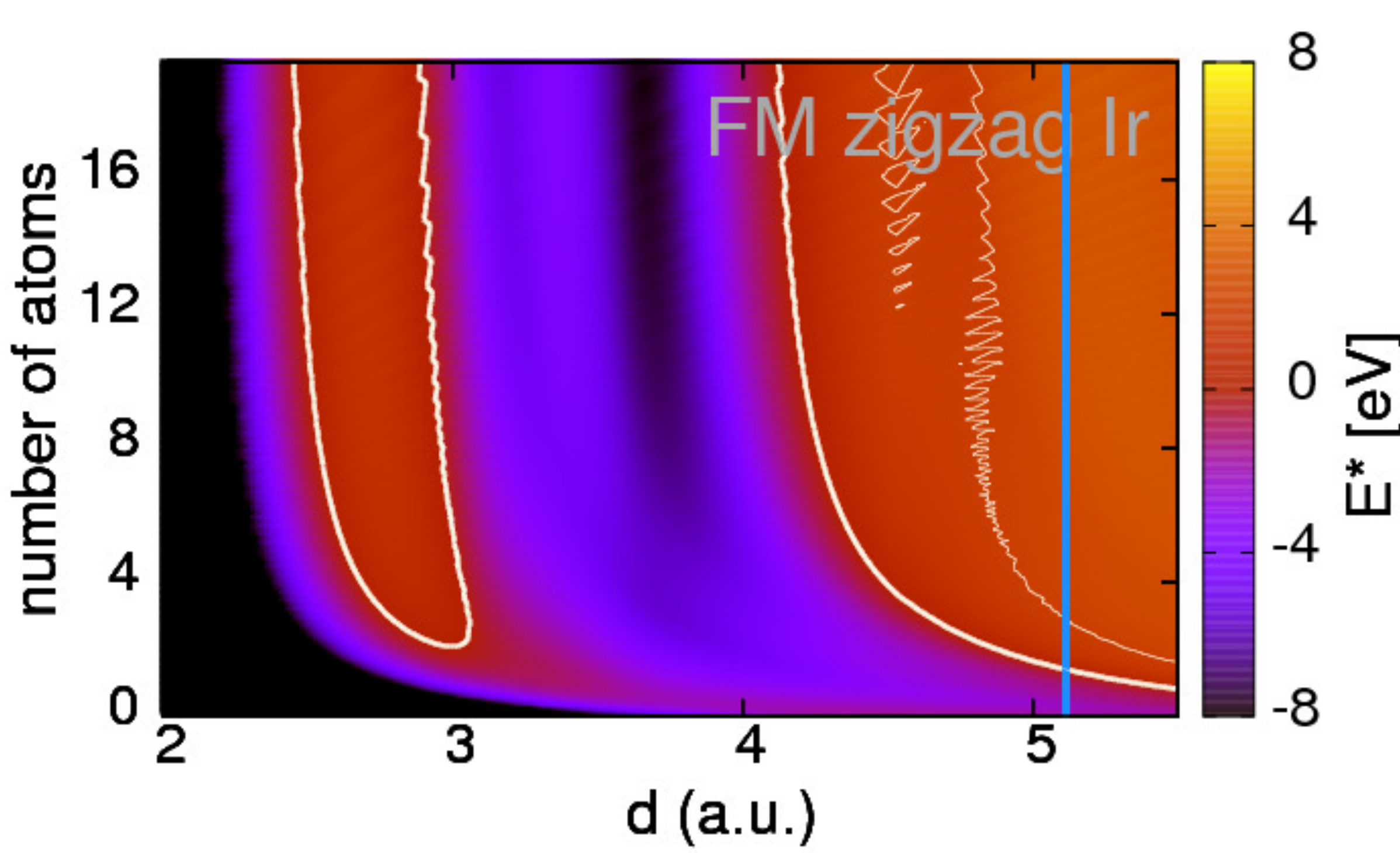} \\
\includegraphics[width=3.00cm, angle=270]{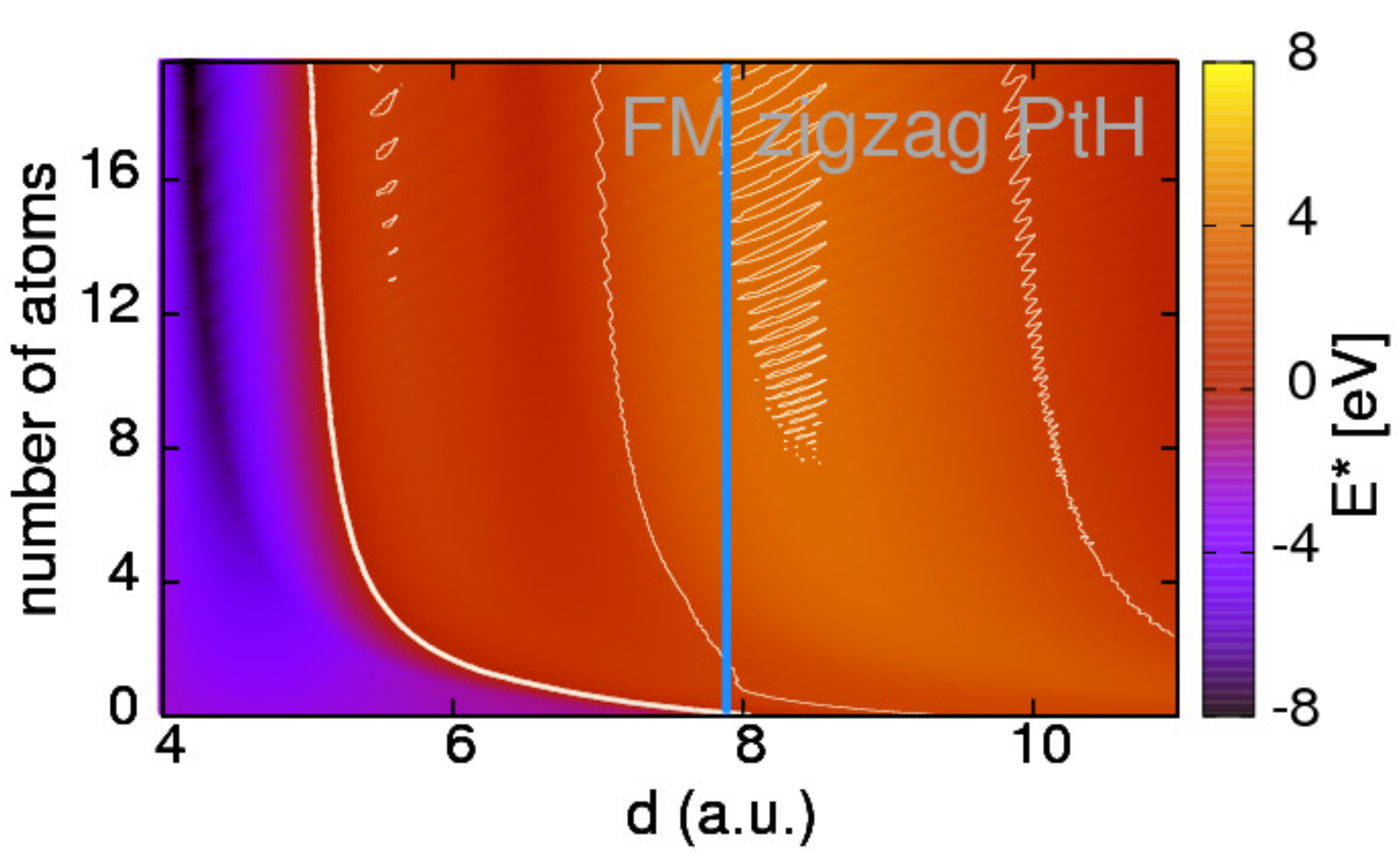}  & & 
\includegraphics[width=3.00cm, angle=270]{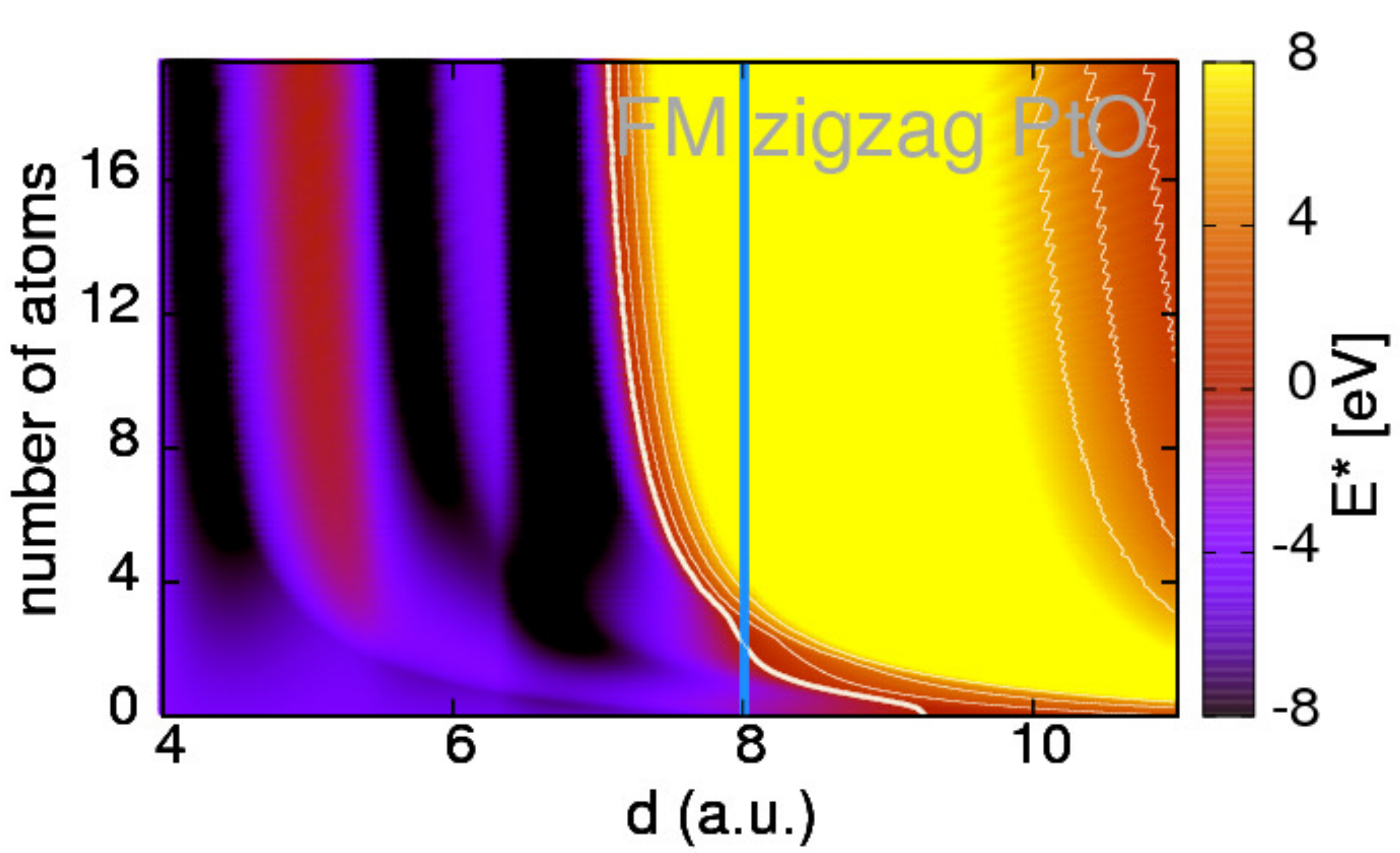} \\ 
\includegraphics[width=3.00cm, angle=270]{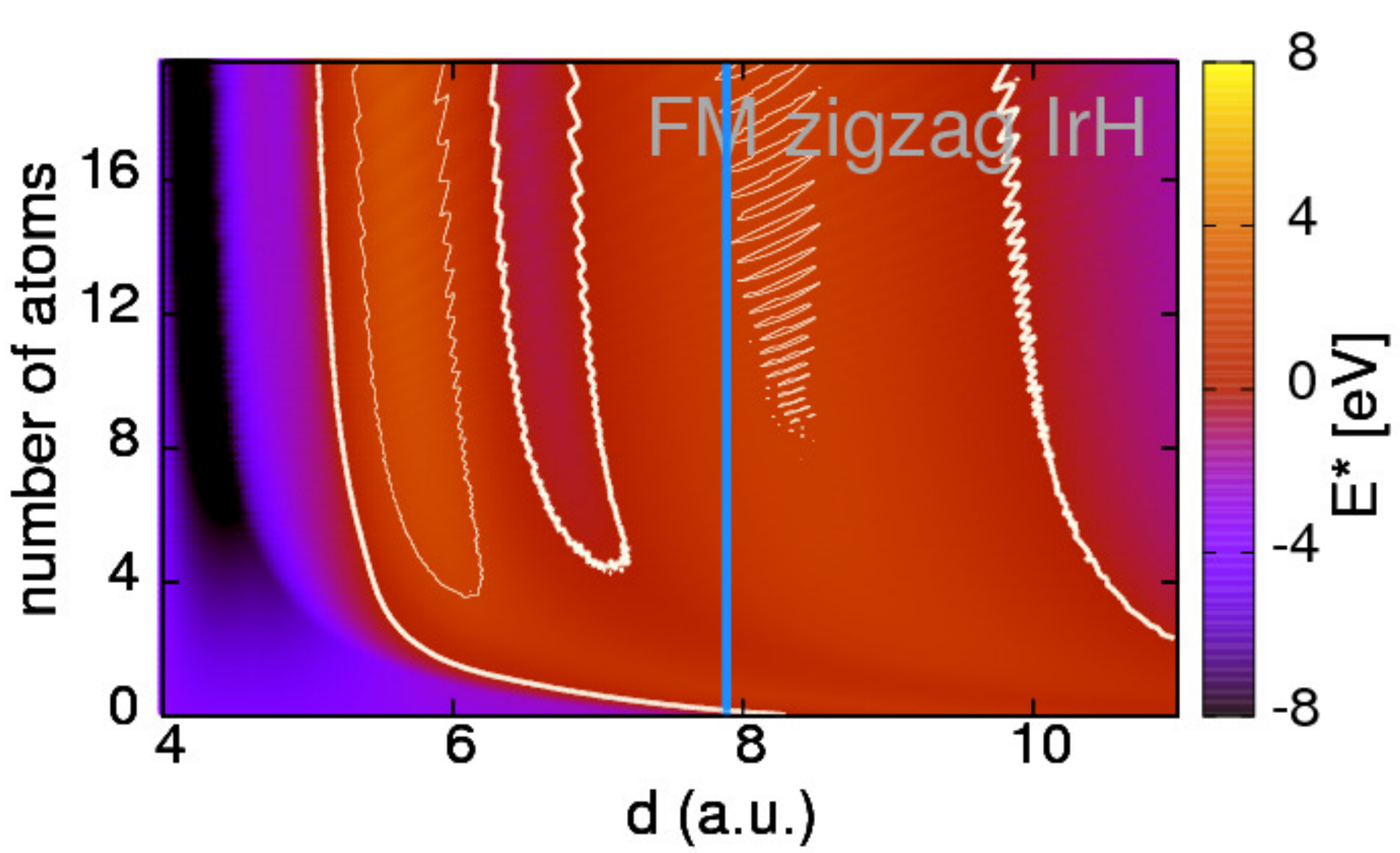}  & & 
\includegraphics[width=3.00cm, angle=270]{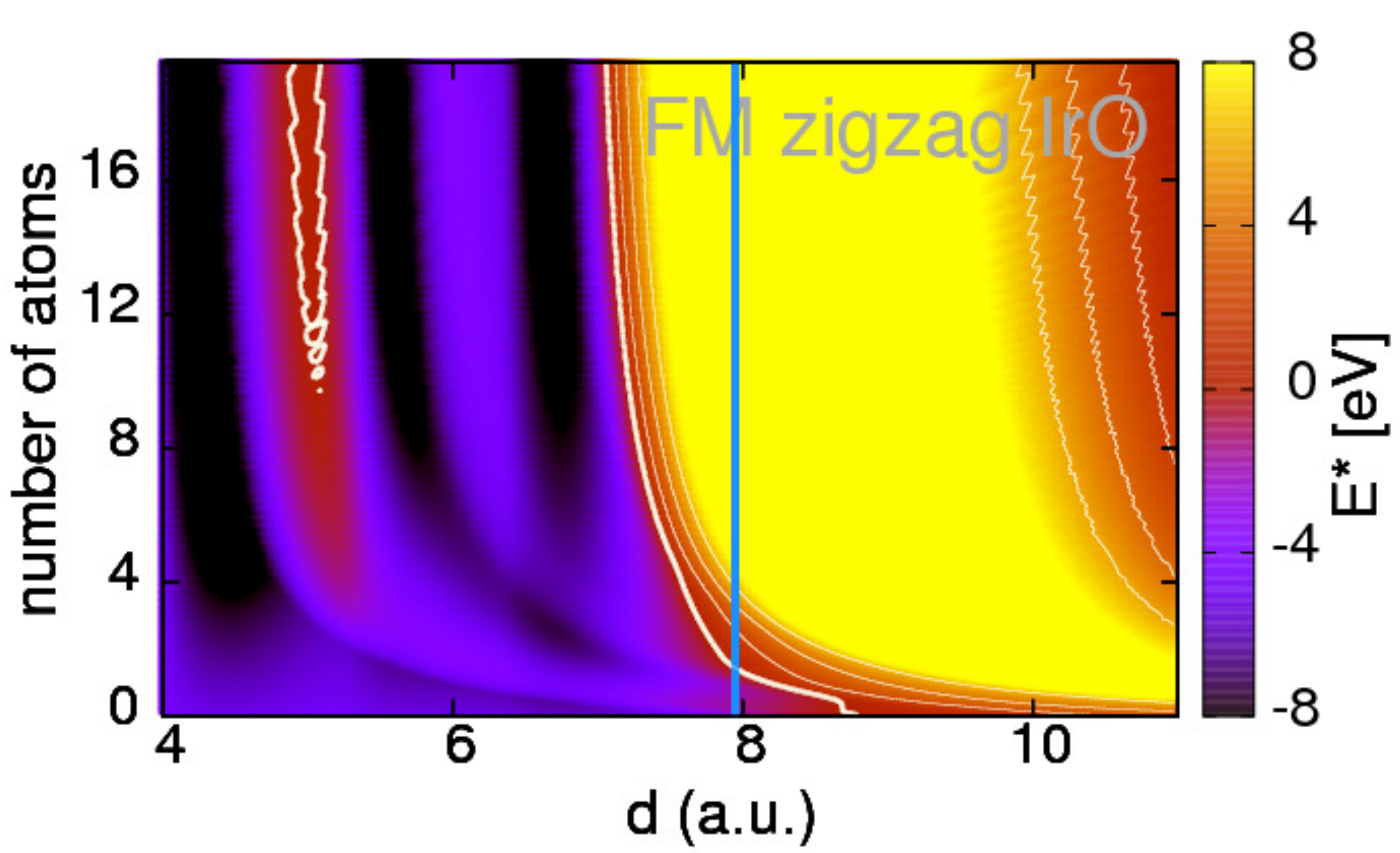} \\ 
\end{tabular}
\end{center}
\caption{Selected energy landscapes $E^*(d,N)$ as a function of interatomic distance ($d$) and 
         number of atoms in the chain ($N$) describing the chain producibility for pure and 
         impurity-assisted Pt and Ir chains, where $E^*$ is defined by the criterion for producibility,
         (eq.~\ref{E*}).
         Thick white lines mark the $E^{*}=0$ isolines indicating regions of producibility ($E^*(d,N) \textgreater 0$). 
         Additional thin isolines are drawn at $E^*(d,N)=2$~eV, 4~eV, 6~eV, and 8~eV in each panel. 
         Note that regions which are outside of the interval [-8~eV, 8~eV] are shown in the same 
         color-coding as the minimal and maximal value of this interval. 
         To rate the stability of chains, for each panel the inflection point $\hat{d}$ 
         in the linear regime of the binding energy potential is depicted by thick blue vertical lines.}
\label{Pcrit}
\end{figure*}

If $E^*(d,N) \textgreater 0$ the P-criterion is satisfied and the chain can increase by one atom for 
that particular pair of parameters $(d,N)$. 
To apply the P criterion to suspended chains, we fit the wire binding energy $\epsilon (d)$ by interpolating the minimum energy curve from the binding energy of several $d_x$ and
$d_z$ distances.\cite{Dinapoli12} Thus the binding energies participating in equation (\ref{E*}) are the ones depicted in Figs.~\ref{Pt} and \ref{Ir}. To analyze the criterion, we
plot the energy $E^*(d,N)$ surface (P-area, in what follows) for several selected cases, as shown in Fig.~\ref{Pcrit}, where we also included the energy landscape of single-atom
Pt linear chains for better comparison.  We do not show the P-areas for all considered magnetic configurations, to avoid confusing the reader with too many plots.  
The area of the phase space where $E^* \textgreater 0$ provides the parameters for which the elongation by one atom (or one pair of atoms when the impurities are considered)
is probable. This probability is greater for higher values of $E^*$, given by the color coding in the figure. 

One of the most distinct differences visible when comparing 
linear and zigzag single-atom chains is that two different areas of positive $E^*$ appear when considering the zigzag geometry. This suggests that in an 
experiment the co-existence of these phases can enhance the elongation properties of the chains by switching from the linear region of producibility into the zigzag one. This 
could be achieved by pushing the tips towards each other, therefore reducing the interatomic distance in the chain that allows the zigzag geometry to be stabilized. This interesting
and useful property is one of the most important outcomes of our extended model, as we are taking into account the relaxation at short interatomic distances into zigzag geometries
which are energetically favorable. It is noteworthy to say that when analyzing the incidence of magnetism and spin orbit interaction (not shown) in the P-area, we do not find
substantial differences, as can be seen from the second and third panels of Fig.~\ref{Pcrit}.

The main differences in the P-areas are obtained when introducing the environmental impurities in the chain formation process 
(see the last two panels of Fig.\ref{Pcrit}). For both $5d$ TMs chains, we can see that when
including H in the chains, the P-area becomes significantly wider, as compared to the pure linear and zigzag chains. This can be attributed to the fact that in TM-H chains there is only one stable zigzag phase with a rather flat minimum,
meaning that the chains are stable for a rather broad range of $\alpha$ angles. Nevertheless, even if with H impurities the chains are producible in a wider range of $d_z$ distances,
the chains will have more probability to grow if it is the O the impurity which assists the chain growth, as it follows from the larger values of $E^*$ (see color coding of 
Fig.\ref{Pcrit}).

To test the stability of the chains, we turn to the string tension, $F(d)$, which is defined as the slope of the binding energy with respect to the distance between the atoms along the $z$-axis, 
i.e. $F(d)=F(d_z)=\delta \epsilon(d_z)/\delta d_z$. The maximum of this quantity, the so-called "break force" $F_0=F(\hat{d})$ can give us a very rough estimate of how far an 
ideal infinite chain can be stretched until it breaks at a distance $\hat{d}$, indicated with thick blue vertical lines in Fig.~\ref{Pcrit}. As we are dealing not only with 
linear monowires but also with planar zigzag wires, in
the energy binding curve we find more than one minimum. In general, the minima located at lower distances $d_z$ are the ones corresponding to zigzag geometries and, 
therefore, the chains are not under high tension in this region. The minima that are located at higher distances correspond to the elongation situation, and so it is in this
region where the chain can be broken if the breaking force is applied. Physically, a successful chain elongation event will occur when the following happens: 
as the chain is stretched, the energy of the system increases up to the point where a lead atom overcomes the chain formation barrier, $\Delta E_{lead}$, and enters the chain. This
reduces the distance in the chain from $d$ to $\tilde d$ and lowers the total energy of an atom in the wire, $E_W(\tilde d) \textless E_W(d)$. The larger the slope of the total
energy $E_W(d)$, the more energy can be gained by relaxing the chain from a distance $d$ to $\tilde d$. Therefore, large values of $F_0$ and small energy barriers $\Delta E_{lead}$
will favor chain elongation.

\begin{table}
\caption{Calculated break forces, $F_0$ (eV/a.u.). The $F_0$ value that corresponds to the easy-axis at the breaking point when including SOC is indicated in bold text.}   
\begin{ruledtabular}
\begin{tabular*}{\hsize}{l@{\extracolsep{0ptplus1fil}}cccc}
Case		&  NM	&  FM	& r-SOC		& z-SOC	\\
\colrule
PtPt            &  1.33	&  1.24	&  1.16 	& \textbf{1.14} \\
PtH             &  1.05	&  1.03	&  1.04		& \textbf{1.03}	\\
PtO             &  2.35	&  2.22	&  \textbf{2.16}&  2.18 	\\
IrIr		&  1.91 &  1.51 &  1.43 	& \textbf{1.37} \\
IrH		&  1.12 &  0.91 &  1.02		& \textbf{1.02} \\
IrO		&  2.69 &  2.38 &  2.38		&  \textbf{2.30} \\
\end{tabular*}
\end{ruledtabular}
\label{F0}
\end{table}

In Table~\ref{F0} we show calculated breaking forces, $F_0$, for the systems we studied. 
As a general trend, we observe that non-magnetic calculations lead to an enhanced stability of the chains, in agreement with previous results~\cite{Alex09, Ferrer07,Ferrer09}.
When comparing the breaking forces for all the studied cases we first note that in pure chains a stronger binding and correspondingly larger values of $F_0$ occur for Ir, and this 
is a direct consequence of the smoother energy profile of Pt chains, discussed in section \ref{results}.  
Introducing $s$-like impurities leads to lower $F_0$ values, contrary to what is obtained for Cu, Ag and Au chains (see Ref.~\onlinecite{Dinapoli12}), where $s$-like impurities help
in strengthening the bonds. This can be attributed to the fact that the presence of H in the chains causes, on one hand, the smallest strengthening of the bonds due to the absence of 
directionality of the $s$-orbitals but, on the other hand, increases the magnetic moments of the metal-atoms thus impeding the formation of 
long chains in break junctions, in agreement whith Ref.~\onlinecite{Alex09}. On the other hand, when the chains are assisted by $p$-like impurities, more pronounced directionality of the
covalent bonds leads to considerably bigger values of the corresponding breaking forces.  

\section{Conclusions}
\label{Conclusions}
In this work we applied the generalization of the chain formation model presented in Ref.~\onlinecite{Dinapoli12} to two late $5d$-TMs chains, Pt and Ir, 
with two prototype impurites, the $s$-like and $p$-like impurities (H and O, respectively). The geometrical extension of the model is essential to describe assisted 
chain creation but also leads to a quantitative refinement of the predicted interchain distances during the chain creation process of the pure chain. 
We have investigated the growth probability and stability of the wires, taking the input from first-principles 
calculations. We find that neither magnetism nor spin-orbit interaction play a crucial role in the structural properties of the atomic chains, but they do 
play a role in their producibility and stability in break junction experiments as the absence of magnetism  leads to an enhanced stability. We also find that
Ir chains are more probable than Pt chains to form and that O-assisted chain growth leads to a strongly enhanced tendency towars chain elongation, when compared
to pure chains. On the other hand, the presence of H-impurities increases the spin magnetic moments of the metal atoms, resulting in a smaller breaking force
and reduced structural stability. 

\acknowledgments{
Y.~M.~gratefully acknowledges funding under the HGF-YIG Programme VH-NG-513 and S.~D.~N acknowledges funding from Conicet, PIP00258.} 

\bibliographystyle{apsrev}
\bibliography{chains}

\end{document}